\documentclass[osid.sty]{article} 

\usepackage[utf8]{inputenc} 

\usepackage{geometry} 
\usepackage{graphicx} 
\usepackage{amsmath}
\usepackage{amssymb}
\usepackage{booktabs} 
\usepackage{array} 
\usepackage{paralist} 
\usepackage{verbatim} 
\usepackage{subfig} 
\usepackage{float}
\usepackage{fancyhdr} 
\usepackage{sectsty}
\allsectionsfont{\sffamily\mdseries\upshape} 
\usepackage[nottoc,notlof,notlot]{tocbibind} 
\usepackage[titles,subfigure]{tocloft} 

\newtheorem{remark}{Remark}

\newcommand{\Tr}{{{\rm Tr}}}

\newcommand{\CI}{{\mathbb{C}}}

\newcommand{\RI}{{\mathbb{R}}}
\newcommand{\id}{{{\rm id}}}

\newcommand{\HI}{{\mathbb{H}}}

\title{A non-Markovian dissipative Maryland model}

\author{F. Benatti\\{\footnotesize\it Department of Physics, University of Trieste, Italy \& INFN, Sezione di Trieste, email: benatti@ts.infn.it}\\[2ex]
        F. Carollo\\{\footnotesize\it Department of Physics, University of Trieste, Italy, email: federico.carollo@ts.infn.it}}

\begin{document}
\date{}
\maketitle

\begin{abstract}
The so-called Maryland model is a linear version of the quantum kicked rotor; it exhibits Anderson localization in momentum space.
By turning the kicks into a Markovian stochastic process, the dynamics becomes a dissipative quantum process described by a discrete family of completely positive maps that allows to explicitly study the relation between divisibility of the maps and the degree of memory of the process.
\end{abstract}

\section{Introduction}
\label{QKMM}

The quantum kicked rotor~\cite{GTOTQKR1} consists of a massive quantum particle moving freely on a circumference and periodically kicked by a cosine pulse that modifies its angular momentum. This kind of dynamics is generated by the following time-dependent singular Hamiltonian
\begin{equation}
\hat{H}_t=\frac{\hat{p}^{2}}{2I}+V_t(\hat{\theta})=\frac{\hat{p}^{2}}{2I}\,+\,K\,\cos\hat{\theta}\sum^{+\infty}_{n=1}\delta(t-n\tau)\ ,
\label{SM1}
\end{equation}
where $K$ is a coupling constant with the dimension of an action, $I$ is the momentum of inertia and $\tau$ the kicking period.

The classical and quantum kicked rotors have been deeply investigated in connection with classical chaos and its signature in quantum systems~\cite{artFishman} and have also been physically implemented~\cite{raizen}. Further interest in this model came from the possibility of being used to investigate the Anderson localization~\cite{artAnderson} and transition from the metallic to the insulating regime~\cite{ATIATDKR}. The quantum kicked rotor localizes the system in angular momentum space
when the period $\tau=4\pi\,\alpha$, with irrational $\alpha$, while in the Anderson model an electron is localized on a $1$-dimensional lattice
in presence of spatial noise. There exists a procedure to map the kicked rotor into an Anderson localization model and vice versa~\cite{artFishman}.

There exists another model, known as quantum linear kicked rotor~\cite{artFishman,EP}, also known as \emph{Maryland Model}, where the kinetic term in the Hamiltonian is taken to be linear in angular momentum,
\begin{equation}
\hat{H}=\hat{p}+V(\hat{\theta})=\hat{p}\,+\,K\,\cos\hat{\theta}\sum^{+\infty}_{n=1}\delta(t-n\tau)\ ,
\label{SM2}
\end{equation}
This model has the advantage of being analytically solvable and can also be mapped into a model exhibiting Anderson localization when $\tau=2\pi\,\alpha$, with irrational $\alpha$~\cite{EP,artTao ma}.

In the following, we shall consider the scenario in which the quantum linear kicked rotor is in contact with an external environment of classical type whose presence is effectively taken into account by treating the kick strengths as stochastic variables. The ensuing decoherence weakens the destructive quantum interferences responsible for the angular momentum localization leading to a diffusive behaviour exhibited by the momentum variance increasing with the square root of time~\cite{Ott,DLS}.
We shall focus upon the kick strengths forming a discrete-time Markov process with varying degree of memory between the realization of the stochastic kick strength at a certain time and the successive one.
The resulting dynamics is described by a discrete family of completely positive maps that compose as a semigroup in absence of memory and are otherwise connected by intertwining maps that are not completely positive as witnessed by the non-monotonic behaviour of the
Hilbert-Schmidt norm. This discrete-time dissipative quantum process appears to be non-Markovian according to
the approach to non-Markovianity through the absence of divisibility~\cite{PiiloManiscalco}--~\cite{vas:PRA:11} (for another approach to non-Markoviaity see~\cite{ChruKoss1} and~\cite{ChruKoss2}). Furthermore, in the present model it can be proved that memory effects make the intertwining maps not even positive.

\section{Maryland Model}

As outlined in the Introduction, we shall focus upon the following Hamiltonian
\begin{equation}
\hat{H}_{x}=\hat{p}+V_{x}(\hat{\theta})=\hat{p}\,+\,\cos\hat{\theta}\sum^{+\infty}_{\ell=1}K(1-x_\ell)\delta(t-\ell\tau)\ ,
\label{SM3}
\end{equation}
where the $x_\ell$ are stochastic variables taking values in $\{0,1\}$ that perturb the otherwise constant kick strength at discrete times $t=\ell\tau$; we shall take them to form a  one-step Markov process $x=\{x_\ell\}$ that we will specify later on.

In the following, we shall frequently use angle representation (for sake of simplicity $\hbar=1$), where
$$
(\hat{p}\psi)(\theta)=-i\psi'(\theta)\ ,\quad
(\cos(\hat{\theta})\psi)(\theta)=\cos\theta\,\psi(\theta)\ ,
$$
and the operator $\hat{p}$ has eigenvectors
$$
\hat{p}\vert\psi_n\rangle\,=\,n\,\vert\psi_n\rangle\ ,\qquad \psi_n(\theta)=
\frac{e^{i\,n\,\theta}}{\sqrt{2\pi}}\ ,\qquad n \in \mathbb{Z}\ .
$$

Because of the Dirac delta, the time-evolutor from $t=(\ell-1)\tau$ and $\ell\tau$ is given by the following unitary operator on the Hilbert space $\displaystyle \mathbb{H}=\mathbb{L}^2([0,2\pi],{\rm d}\theta)$ of the system:
\begin{equation}
\hat{U}_{\ell\tau}=\hat{V_\ell}\,e^{-i\hat{p}\tau}\ ,\quad \hat{V_\ell}=e^{-iK(1-x_\ell)\,\cos\hat{\theta}}\ .
\label{SM4}
\end{equation}
Let us consider an initial state, namely a density matrix  $\hat{\rho}$ acting on $\mathbb{H}$; the state $\hat{\rho}_N$ at discrete time $t=N\tau$ is thus obtained as
\begin{equation}
\hat{\rho}_N=\hat{U}_{N}\,\hat{\rho}\, U^\dag_N\ ,\quad U_N=\hat{U}_{N\tau}\hat{U}_{(N-1)\tau}\hat{U}_{(N-2)\tau}\dots\hat{U}_{\tau}\ .
\label{SM5}
\end{equation}
Using that
\begin{equation}
e^{i\hat{p}\,\ell\tau}e^{-iK(1-x_\ell)\cos\hat{\theta}}e^{-i\hat{p}\,\ell\tau}=e^{-iK(1-x_\ell)\cos(\hat{\theta}+\ell\tau)}\ ,
\label{SM9}
\end{equation}
one rewrites
\begin{equation}
\hat{U}_{N}=e^{-i\,\hat{p}\,N\,\tau}\,e^{-i\sum_{\ell=1}^{N}K(1-x_\ell)\cos(\hat{\theta}+\ell\tau)}\ .
\label{SM10}
\end{equation}
The variance of the angular momentum with respect to a time-evolving state initially localized in momentum space is a measure of how much the latter spreads.
In the Heisenberg picture 
\begin{eqnarray}
\hat{p}_{N}&=&\hat{U}_{N}^{\dagger}\,\hat{p}\,\hat{U}_{N}=e^{+iK\sum_{\ell=1}^{N}(1-x_\ell)\cos(\hat{\theta}+\ell\tau)}\,\hat{p}\,
e^{-iK\sum_{\ell=1}^{N}(1-x_\ell)\cos(\hat{\theta}+\ell\tau)}
\label{SM11}\\
&=&\hat{p}+\sum_{\ell=1}^{N}K(1-x_\ell)\sin(\hat{\theta}+\ell\tau)\ .
\label{SM23}
\end{eqnarray}
Setting $f_N(\hat{\theta})=\sum_{\ell=1}^{N}K(1-x_\ell)\sin(\hat{\theta}+\ell\tau)$ and choosing as initial state the zero-angular momentum
eigenstate $\vert 0\rangle$ of $\hat{p}$, one gets
\begin{eqnarray}
\langle 0\vert\,\hat{p}_N\,\vert 0\rangle&=&\langle 0\vert\,f_N(\hat{\theta})\,\vert 0\rangle=K\sum_{\ell=1}^{N}(1-x_\ell)\frac{1}{2\pi}\int_0^{2\pi}{\rm d}\theta\, \sin(\theta+\ell\tau)=0
\label{SM26}\\
\nonumber
\Delta_{\vert 0\rangle}^{2}\hat{p}_N&=&\langle0\vert\hat{p}^{2}_N\vert 0\rangle-\Big(\langle0\vert\hat{p}_{N}\vert 0\rangle\Big)^2=\langle0|f_N^{2}(\hat{\theta})|0\rangle\\
&=&K^2\sum_{j,k=1}^{N}\frac{(1-x_j)(1-x_k)}{2}\,\cos[(j-k)\tau]\ .
\label{SM27}
\end{eqnarray}
As this result depends on the explicit realization of the stochastic process up to time $t=N\tau$, in order to get a physically testable quantity one has to average over all possible realizations.

\begin{figure}[H]
\begin{center}
\includegraphics[width=0.70\textwidth]{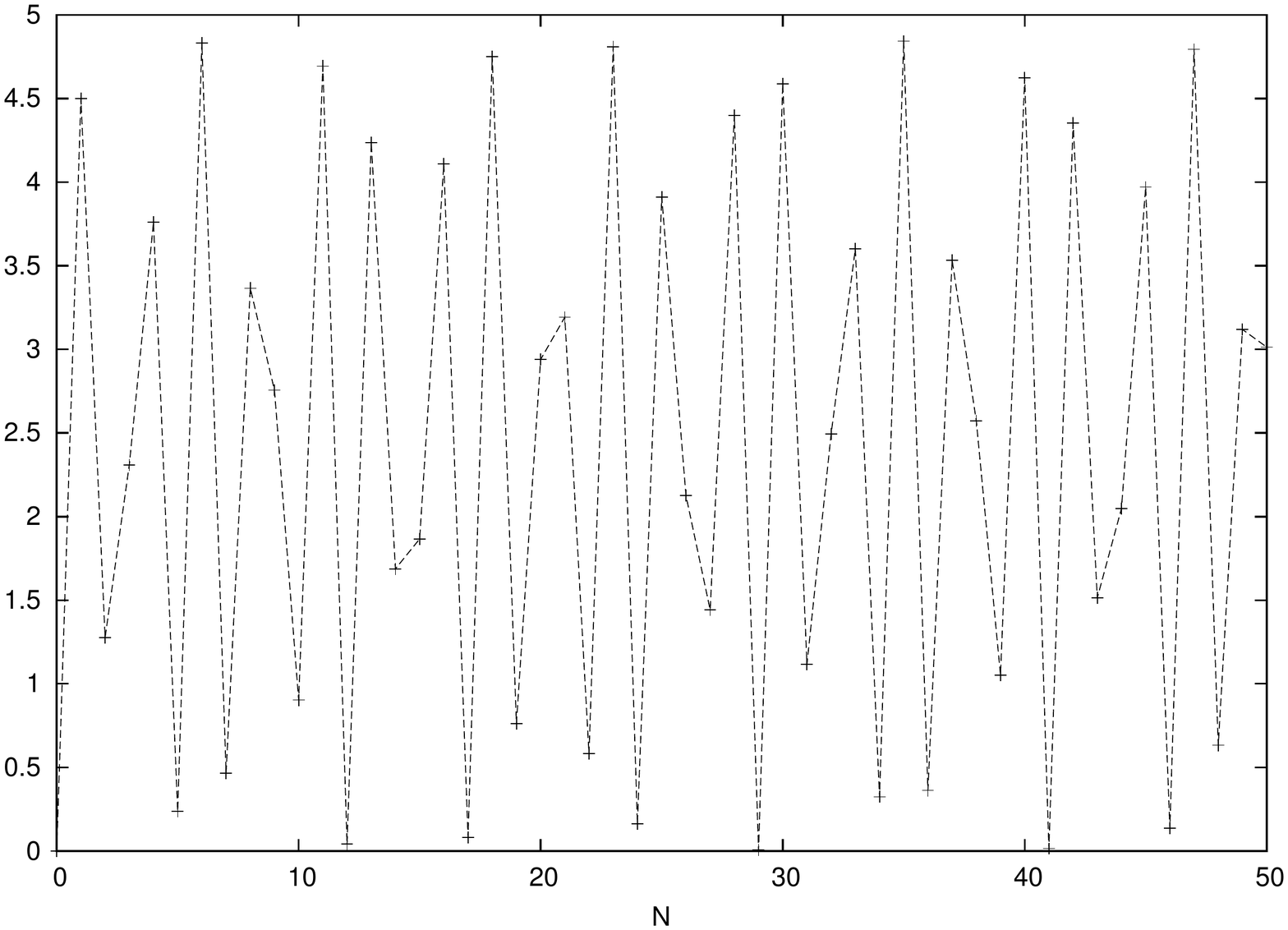}
\caption{$\langle\Delta^2_{\vert0\rangle}\hat{p}\rangle$ as function of $N\tau$ for $K=3$ and $\tau={2\pi}\sqrt{2}$ without stochastic kicks.}
\label{IM0}
\end{center}
\end{figure}

\begin{figure}[H]
\begin{center}
\includegraphics[width=0.70\textwidth]{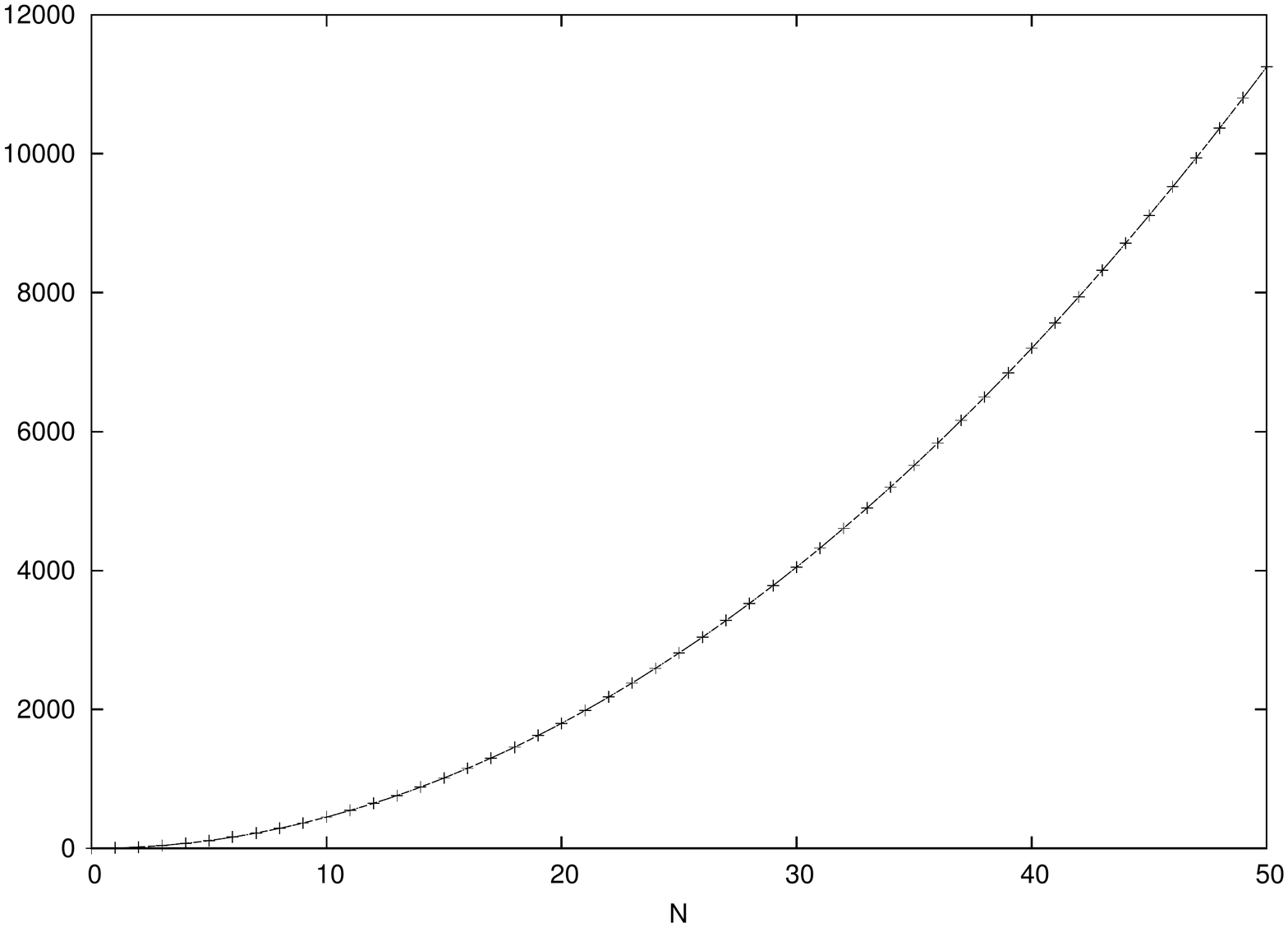}
\caption{$\langle\Delta^2_{\vert0\rangle}\hat{p}\rangle$ as function of $N\tau$ for $K=3$ and $\tau={2\pi}$ without stochastic kicks.}
\label{IM0a}
\end{center}
\end{figure}

Before going to that, let us consider the case when the kick strengths are not stochastically perturbed, namely $x_\ell=0$ for all $\ell$; then, the variance is a bounded function of $N$  when $\tau=2\pi\,\alpha$ with $\alpha$ irrational, with corresponding localization, that is there is no spreading in angular momentum as illustrated in Fig.~\ref{IM0}.
On the other hand, Fig.~\ref{IM0a} shows the increasing of the angular momentum variance for $\tau=2\pi$.

From now on, the attention will be focused on the case $\tau=2\pi\,\alpha$ with $\alpha$ irrational.
As already stated, we take as $x=\{x_\ell\}$ a discrete $1$-step Markov process with each $x_\ell$ taking the values $0,1$ with probabilities $p_0=p_1=1/2$.
The stochastic strengths are chosen in such a way that, at each tick of time of period $\tau$, the rotor is kicked or not with equal probability.
Furthermore, the probability for the realization of the stochastic process $\bar{x}_N=(x_1,x_2,\ldots,x_N)$ will be denoted by
$p(\bar{x}_N)$ of the form
\begin{equation}
\label{Markov}
p(\bar{x}_N)=T_{x_Nx_{N-1}}\,T_{x_{N-1}x_{N-2}}\cdots T_{x_2x_1}\,p_{x_1}\ , \quad x_j\in\{0,1\}\ ,
\end{equation}
with transition coefficients $T_{x_{i+1}x_i}$ for the value $x_i$ be followed by $x_{i+1}$, given by
\begin{equation}
\label{TP}
T_{x_{i+1}x_i}=a\delta_{x_{i+1}x_i}+(1-a)p(x_{i+1})\ ,\quad x_{i+1},x_i=0,1\ ,
\end{equation}
so that the $2\times 2$ transition matrix $T=[T_{x_{i+1}x_i}]$ reads
\begin{equation}
\label{TM}
T=a\,\begin{pmatrix}1&0\cr0&1\end{pmatrix}\,+\,\frac{1-a}{2}\,\begin{pmatrix}1&1\cr 1&1\end{pmatrix}\ ,\quad 0\leq a\leq 1\ .
\end{equation}
The first contribution to the transition matrix represents a Markov process with full memory, in the sense that only the first kick strength is stochastic and if it takes the value $K$ $(0)$, it will keep it $K$ $(0)$ at all other times. On the other hand, the second contribution to the transition matrix gives rise to a Bernoulli process with no memory, that is successive kick strengths are independent of one another. Therefore, by increasing $a$ from $0$ to $1$ one interpolates between a Markov process with no memory and one with full memory.

The form of the transition matrix is particularly simple and yields
\begin{equation}
\label{TMn}
T^n=a^n\,\begin{pmatrix}1&0\cr0&1\end{pmatrix}\,+\,\frac{1-a^n}{2}\,\begin{pmatrix}1&1\cr 1&1\end{pmatrix}\ ,\quad 0\leq a\leq 1\ .
\end{equation}
Furthermore, the probability vector $\vert \pi\rangle=1/2(1,1)^T$ is left invariant by the transition matrix $T$ so that the Markov process is stationary:
\begin{equation}
\label{stationary}
T\vert\pi\rangle=\vert\pi\rangle\,\Longrightarrow \sum_{x_1=0}^1p(x_1,x_2)=p(x_2)\ .
\end{equation}
The characteristic function of such a process reads
\begin{eqnarray}
\label{charfunctmarkov}
\Lambda_N(\bar{u}_N)&=&\sum_{\bar{x}_N\in\{0,1\}^N}\,e^{-i\langle\bar{x}_N\vert\bar{u}_N\rangle}\,p(\bar{x}_N)=
(1,1)\,T_{\bar{u}_N}\,\frac{1}{2}\begin{pmatrix}1\cr e^{-iu_1}
\end{pmatrix}\\
T_{\bar{u}_N}&=& T_{u_N}T_{u_{N-1}}\cdots T_{u_2}\ ,\quad
\label{charfunctmarkov2}
T_u=E_u\,T\ ,\quad E_u=\begin{pmatrix}1&0\cr 0&e^{-iu}\end{pmatrix}\ .
\end{eqnarray}

\begin{figure}[H]
\begin{center}
\includegraphics[width=0.7\textwidth]{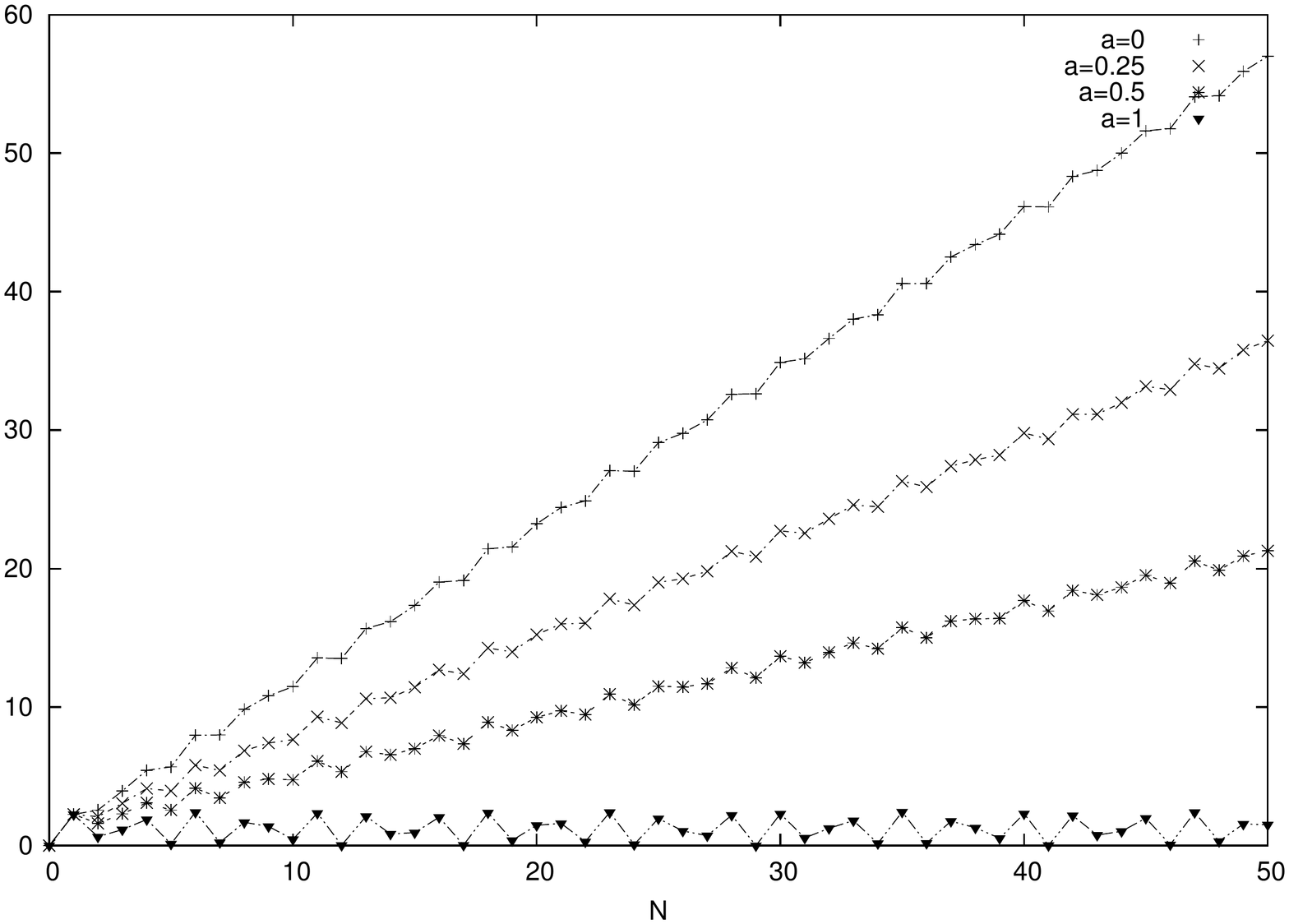}
\caption{$\langle\Delta^2_{\vert0\rangle}\hat{p}\rangle$ as function of $N\tau$ for the Markov process with $K=3$ and $\tau={2\pi}\sqrt{2}$.}
\label{IMM1}
\end{center}
\end{figure}

We shall denote by $\langle\,\cdot\rangle$ the averages with respect to the probability distribution; then, the first moments $\langle x_\ell\rangle_N=1/2$ while the second moments $\langle x_j\,x_k\rangle_N$ can be computed as follows: suppose $k\geq j$, then
$$
\langle x_j\,x_k\rangle_N=\sum_{x_j,\cdots,x_k}x_j\,x_k\sum_{x_N}\Big(T^{N-k}\Big)_{x_Nx_k}\,\Big(T^{k-j}\Big)_{x_kx_j} \sum_{x_1}\Big(T^{j-1}\Big)_{x_jx_1}p(x_1)\ .
$$
From the stationarity of the process~(\ref{stationary}) and~(\ref{TMn}) one gets
\begin{equation}
\label{2ndmomMark}
\langle x_j\,x_k\rangle_N=\frac{1}{4}\Big(1+a^{k-j}\Big)\ .
\end{equation}
Then, the momentum variance with respect to the angular momentum eigenstate $\vert0\rangle$ reads
\begin{eqnarray}
\langle\Delta^{2}_{\vert0\rangle}\hat{p}_N\rangle
=\sum_{j,k=1}^{N}\frac{K^2}{8}\Big(1+a^{k-j}\Big)\,\cos[(j-k)\tau]\ .
\label{SM31markov}
\end{eqnarray}
Figure~\ref{IMM1} shows a diffusive behaviour which tends to disappear as $a$ approaches $1$.

In the case of $a=1$, with probability $1/2$ there are no kicks and with the same probability there are always kicks of same strength $K$; therefore, the full memory case corresponds to the deterministic Maryland model (divided by two as this is the probability for a process with kicks) as showed in the following figure for irrational $\tau/(2\pi)$.

\begin{figure}[H]
\begin{center}
\includegraphics[width=0.70\textwidth]{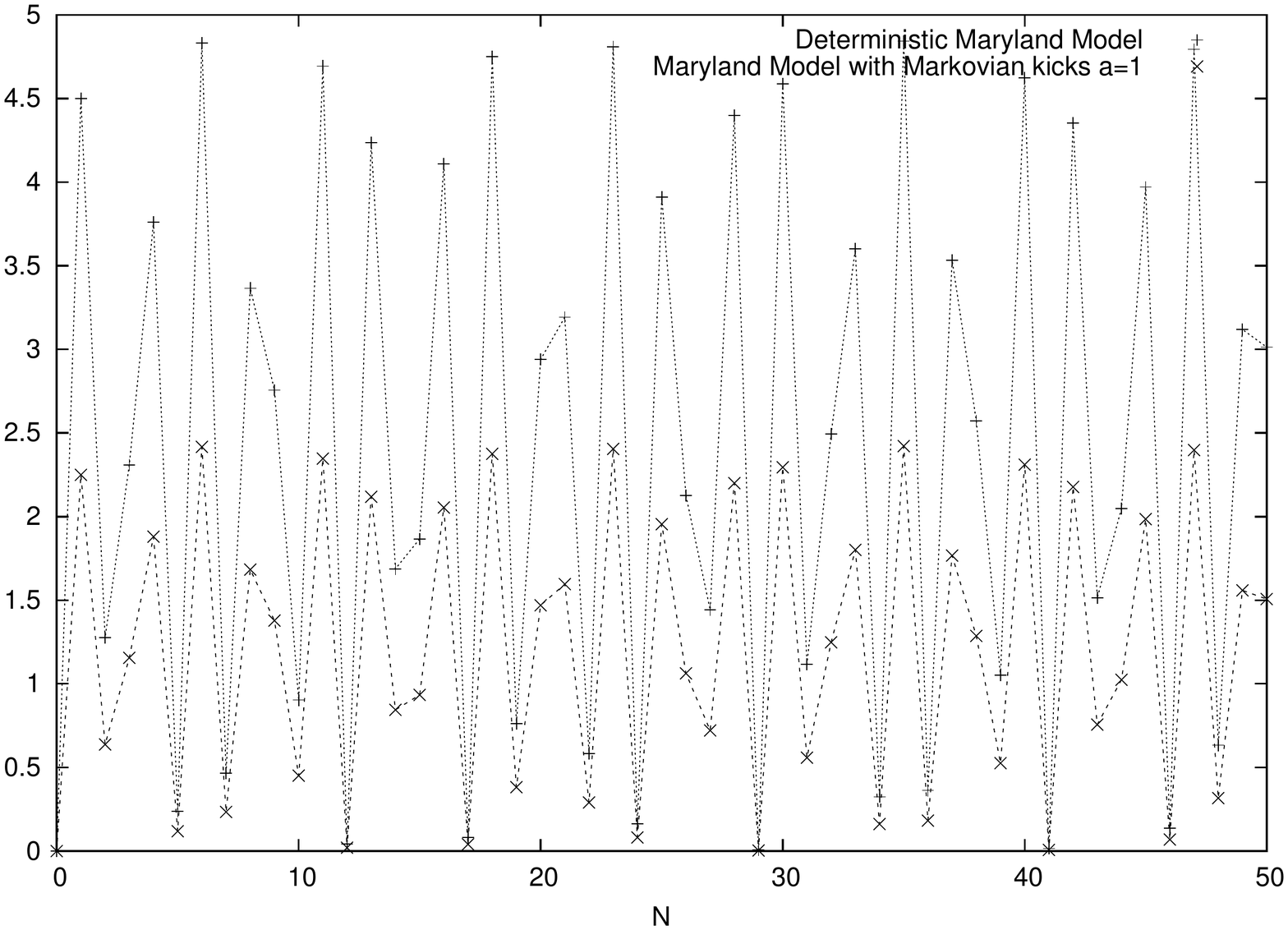}
\caption{Comparison of momentum variances $\Delta_{\vert0\rangle}\hat{p}$ as function of $N\tau$ in the deterministic Maryland model and in the stochastic Markov model with with full memory: $K=3$, $\tau={2\pi}\sqrt{2}$ and $a=1$.}
\label{IMM2irr}
\end{center}
\end{figure}

\section{Dynamical maps and non-Markovianity}

Like for the physical variance, a physical time-evolution for the density matrix of the linear kicked rotor with stochastic kicks is obtained from~(\ref{SM5})
by averaging with respect to the realizations $\bar{x}_{N}=(x_1,x_2,\ldots,x_N)$ of the stochastic process up to time $t=N\tau$. The averaging process yields the following maps on the density matrices of the quantum linear kicked rotor:
\begin{equation}
\hat{\rho}\mapsto \Phi_{N}[\hat{\rho}]=\sum_{\bar{x}_N\in\{0,1\}^N} p(\bar{x}_N)\,\prod_{j=1}^N\hat{U}_{j\tau}(x_j)\,\hat{\rho}\,
\prod_{j=1}^N\hat{U}^\dag_{j\tau}(x_j)\ ,
\label{SM32}
\end{equation}
where the unitary operators $\hat{U}_{j\tau}(x_j)$ are as in~(\ref{SM4}) and depend on the realization of the stochastic process at $t=j\tau$,
$p(\bar{x}_{N})$ is the probability distribution (\ref{Markov}).
The maps $\Phi_N$ are unital and trace-preserving:
\begin{eqnarray}
\label{unital0}
\Phi_{N}[1]&=&\sum_{\bar{x}_N\in\{0,1\}^N} p(\bar{x}_N)\,\prod_{j=1}^N\hat{U}_{j\tau}(x_j)\,1\,
\prod_{j=1}^N\hat{U}^\dag_{j\tau}(x_j)=\sum_{\bar{x}_N\in\{0,1\}^N} p(\bar{x}_N)\,=\, 1\\
\nonumber
\Tr(\Phi_N[\hat{\rho}])&=&\sum_{\bar{x}_N\in\{0,1\}^N} p(\bar{x}_N)\,\Tr\left[\prod_{j=1}^N\hat{U}_{j\tau}(x_j)\right]\,\hat{\rho}\,
\left[\prod_{j=1}^N\hat{U}^\dag_{j\tau}(x_j)\right]=\sum_{\bar{x}_N\in\{0,1\}^N} p(\bar{x}_N)\,\Tr(\hat{\rho})\\
\label{trace0}
&=&\,\Tr(\hat{\rho})\ .
\end{eqnarray}
\medskip

\begin{remark}
\label{rem1}
The dynamical maps $\Phi_N$ evolve any given initial density matrix into a density matrix at time $t=N\tau$; indeed, because of the positivity of the probability measure, they are a convex combination of unitary actions and thus a (generalized) Kraus-Stinespring form of completely positive maps on the space of states~\cite{AlickiFannes}. Therefore, they preserve the positivity not only of $\hat{\rho}$, but the composite maps $\Phi_N\otimes\id_n$ also preserve the positivity of generic density matrices acting on the Hilbert spaces $\mathbb{H}\otimes\mathbb{C}^n$, for all $n\geq 1$.
This latter fact is essential for the physical consistency of the maps as dynamical maps; indeed, one could always couple the quantum linear kicked rotor with an $n$-level system in such a way that the states of the coupled system evolve according to $\Phi_N\otimes\id_n$: if $\Phi_N$ were not completely positive, one could always find an $n$ and an entangled initial state $\hat{\rho}_{ent}$ of the compound system such that $\Phi_N\otimes\id_n[\hat{\rho}_{ent}]$ develops for some $N$ negative eigenvalues then losing positivity and thus its physical interpretation as a state.
\end{remark}
\medskip

If the stochastic process $x$ consisted of independent identically distributed stochastic variables, then
$p(\bar{x}_N)=\prod_{j=1}^Np(x_j)$ and $\Phi_N=(\Phi)^N=\Phi\circ\Phi\circ\cdots\Phi$, $N$ times, where
\begin{equation}
\Phi[\hat{\rho}]=\sum_{x=0,1}\,e^{-iK(1-x)\cos\hat{\theta}}\,e^{-i\hat{p}\tau}\,\hat{\rho}\, e^{i\hat{p}\tau}\,e^{iK(1-x)\cos\hat{\theta}}\ .
\label{SM46}
\end{equation}
Independence of $x_l$ means the absence of memory among realizations of the noise at different times which is reflected by the fact that the
family of dynamical maps is a discrete semigroup: $\Phi_N=\Phi\circ\Phi_{N-1}=\Phi_{N-1}\circ\Phi$.
The presence of memory between kicks at different times spoils the semigroup property, but one can still wonder whether there exist intertwining maps
$\Phi_{N,M}$ on the space of states such that
\begin{equation}
\label{intertwining}
\Phi_N=\Phi_{N,M}\circ\Phi_M\ ,\qquad 1\leq M\leq N\ ,\ \hbox{then}\quad \Phi_{N,M}=\Phi_N\circ\Phi_M^{-1}\ ,
\end{equation}
so that the issue at stake is to invert maps as in~(\ref{SM32}).

Let us introduce the real $N$-dimensional vectors $\bar{I}_N=(1,1,\cdots,1)$,
\begin{eqnarray*}
\bar{J}_N(\bar{x}_N)&=&\bar{I}-\bar{x}_N=(1-x_1,1-x_2,\ldots,1-x_N)\\
\bar{\nu}_{N}(\hat{\theta})&=&K(\cos(\hat{\theta}+\tau),\cos(\hat{\theta}+2\tau)\,\ldots,\cos(\hat{\theta}+N\tau))
\end{eqnarray*}
and denote by $\langle\cdot\,\vert\,\cdot\rangle$ the scalar products of such vectors; then, using~(\ref{SM10}), one rewrites
\begin{eqnarray}
\label{SM33}
\Phi_{N}[\hat{\rho}]&=&e^{-i\hat{p}\,N\,\tau}\,\Psi_N[\hat{\rho}]\,e^{i\hat{p}\,N\,\tau}\\
\Psi_N[\hat{\rho}]&=&\sum_{\bar{x}_N\in\{0,1\}^N} p(\bar{x}_N)\,e^{-i\langle\bar{J}_N(\bar{x}_{N})\vert\bar{\nu}_{N}(\hat{\theta})\rangle}\,\hat{\rho}\,
e^{i\langle\bar{J}_N(\bar{x}_{N})\vert\bar{\nu}_{N}(\hat{\theta})\rangle}\ .
\label{SM33a}
\end{eqnarray}
In order to find the inverse of the map $\Psi_N$, one can proceed as follows: from
\begin{equation}
\langle \theta_{1}\vert\Psi_N[\hat{\rho}]\vert\theta_{2}\rangle\,=\,\langle\theta_{1}\vert\hat{\rho}\vert\theta_{2}\rangle\,\sum_{\bar{x}_N\in\{0,1\}^N} p(\bar{x}_N)\, e^{i\langle\bar{J}_N(\bar{x}_{N})\vert\bar{\nu}_{N}(\theta_{2})-\bar{\nu}_{N}(\theta_{1})\rangle}\ ,
\label{SM67}
\end{equation}
one can formally invert the expression, yielding
\begin{equation}
\langle \theta_{1}\vert\hat{\rho}\vert\theta_{2}\rangle\,=\,{\rm e}^{-i\langle \bar{I}_N\vert\bar{\nu}_{N}(\theta_{2})-\bar{\nu}_{N}(\theta_{1}\rangle}\,\frac{\langle \theta_{1}\vert\Psi_N[\hat{\rho}]\vert\theta_{2}\rangle}
{\sum_{\bar{x}_N\in\{0,1\}^N} p(\bar{x}_N)\,e^{-i\langle\bar{x}_{N}|\bar{\nu}_{N}(\theta_{2})-\bar{\nu}_{N}(\theta_{1})\rangle}}\ .
\label{SM68}
\end{equation}
By means of a $N$-dimensional Dirac delta
$$
\delta\left(\bar{u}_{N}-\bar{\nu}_{N}(\theta_{2})+\bar{\nu}_{N}(\theta_{1})\right)
=\int_{\RI^N} \frac{{\rm d}\bar{v}_{N}}{(2\pi)^{N}}\, e^{i\langle\bar{v}_{N}\vert\bar{u}_{N}-\bar{\nu}_{N}(\theta_{2})+\bar{\nu}_{N}(\theta_{1})\rangle}\ ,
$$
used to represent the fraction, one can write the representation independent operator expression
\begin{equation}
\hat{\rho}=\int \frac{{\rm d}\bar{u}_{N}d\bar{v}_{N}}{(2\pi)^{N}}\,\frac{e^{i\langle\bar{v}_{N}\vert\bar{u}_{N}\rangle}}{\Lambda_N(\bar{u}_N)}\, e^{i\langle \bar{I}_N+\bar{v}_{N}\vert\bar{\nu}_{N}(\hat{\theta})\rangle}\,\Psi_N[\hat{\rho}]\, 
e^{-i\langle \bar{I}_N+\bar{v}_{N}\vert\bar{\nu}_{N}(\hat{\theta})\rangle}\ ,
\label{SM71}
\end{equation}
where $\Lambda_N(\bar{u}_N)$ is the characteristic function of the probability distribution over the noise-realizations $\bar{x}_N$.
Therefore, the formal inverse of the map $\Phi_N$ in~(\ref{SM32}) is the following linear map $\Phi_N^{-1}$ acting on the operators $\hat{X}$ on $\HI$,
\begin{equation}
\label{inverse}
\Phi_N^{-1}[\hat{X}]=\int \frac{{\rm d}\bar{u}_{N}{\rm d}\bar{v}_{N}}{(2\pi)^{N}}\,\frac{e^{i\langle\bar{v}_{N}\vert\bar{u}_{N}\rangle}}{\Lambda_N(\bar{u}_N)}\, e^{i\langle \bar{I}_{N}+\bar{v}_{N}\vert\bar{\nu}_{N}(\hat{\theta})\rangle}\,e^{i\hat{p}\,N\,\tau}\,\hat{X}\,e^{-i\hat{p}\,N\,\tau}\, e^{-i\langle \bar{I}_{N}+\bar{v}_{N}\vert\bar{\nu}_{N}(\hat{\theta})\rangle}\ .
\end{equation}
Let
\begin{equation}
\label{invchar}
\tilde{\Lambda}(\bar{v}_N)=\int \frac{{\rm d}\bar{u}_{N}}{(2\pi)^{N}}\,\frac{e^{i\langle\bar{v}_{N}\vert\bar{u}_{N}\rangle}}{\Lambda_N(\bar{u}_N)}\ ,
\end{equation}
be the Fourier transform of the inverse of the characteristic function; one rewrites
\begin{equation}
\label{inverse1}
\Phi_N^{-1}[\hat{X}]=\int{\rm d}\bar{v}_N\, \tilde{\Lambda}_N(\bar{v}_N)\,e^{i\langle \bar{I}_{N}+\bar{v}_{N}\vert\bar{\nu}_{N}(\hat{\theta})\rangle} \,e^{i\hat{p}\,N\,\tau}\,\hat{X}\,e^{-i\hat{p}\,N\,\tau}\,e^{-i\langle \bar{I}_{N}+\bar{v}_{N}\vert\bar{\nu}_{N}(\hat{\theta})\rangle}\ .
\end{equation}
Like~(\ref{SM32}), the expression above resembles a continuous Kraus-Stinespring representation of a completely positive map; however, unlike the Fourier transform of the characteristic function $\Lambda_N(\bar{u}_N)$ which yields the probability distribution and is thus positive, $\tilde{\Lambda}_N(\bar{v}_M)$ need not be positive, whence $\Phi_N^{-1}$ and thus $\phi_{N,M}=\Phi_N\circ\Phi_M^{-1}$ need not be completely positive.

\begin{remark}
\label{NCPrem}
The Fourier transform $\tilde{\Lambda}_N(\bar{v}_N)$ need not exist as a function.
However, since it appears within an integration with respect to $\bar{v}_N$ it can always be interpreted as a distribution over a suitable class of test functions. These integration functions immediately appear when computing quantities as
$$
\langle\varphi\vert\Phi_N^{-1}[\hat{X}]\vert\psi\rangle=\int{\rm d}\bar{v}_N\, \tilde{\Lambda}_N(\bar{v}_N)\, \langle\varphi\vert\hat{V}_N(\bar{v}_N) \,e^{i\hat{p}\,N\,\tau}\,\hat{X}\,e^{-i\hat{p}\,N\,\tau}\, \hat{V}^\dag_N(\bar{v}_N)\vert\psi\rangle\ ,
$$
where the operators are as in~(\ref{inverse1}), with suitable choices of vectors $\vert\varphi\rangle\,,\,\vert\psi\rangle\in\HI$.
Notice that $\Phi_N^{-1}$ is unital and trace preserving, namely
\begin{eqnarray}
\label{unital}
\Phi_N^{-1}[1]&=&\int{\rm d}\bar{v}_N\, \tilde{\Lambda}_N(\bar{v}_N)\, 1\, =\, 1\\
\nonumber
\Tr(\Phi^{-1}_N[\hat{\rho}])&=&\int{\rm d}\bar{v}_N\, \tilde{\Lambda}_N(\bar{v}_N)\, \Tr\Big(\hat{V}_N(\bar{v}_N) \,e^{i\hat{p}\,N\,\tau}\,\hat{\rho}\,e^{-i\hat{p}\,N\,\tau}\, \hat{V}^\dag_N(\bar{v}_N)\Big)\\
\label{trace}
&=&\int{\rm d}\bar{v}_N\, \tilde{\Lambda}_N(\bar{v}_N)\, \Tr(\hat{\rho})=\Tr(\hat{\rho})\ .
\end{eqnarray}
\end{remark}

In the case of a continuous-time family of completely positive maps $\{\Phi_t\}_{t\geq 0}$, while Markovianity is identified by the composition law
\begin{equation}
\label{Markovianity}
\Phi_t\circ\Phi_s=\Phi_s\circ\Phi_t=\Phi_{s+t}\ ,\qquad\forall\,s,t\geq 0\ ,
\end{equation}
a criterion of non-Markovianity is taken to be the indivisibility of the
one-parameter family of maps $\Phi_t$~(\ref{SM32}), namely the absence of completely positive intertwining maps $\Phi_{t,s}$ such that
\begin{equation}
\label{non-Markovianity}
\Phi_t=\Phi_{t,s}\circ\Phi_s \qquad \forall\, s,t\geq 0\ .
\end{equation}
The absence of completely positive intertwining maps can be witnessed by the non-monotonic behavior of quantities that, under completely positive maps, would either be always non-decreasing or non-increasing.
Consider a function $F$ on the space of states which is monotonically non-increasing under completely positive maps $\Psi$, that is $F\circ\Psi\leq F$. Since the maps $\Phi_t$ are completely positive, it follows that $F_t=F\circ\Phi_t\leq F$ for all $t$; however, if $\Phi_{t,s}$ is not completely positive, then $F_t=F\circ\Phi_{t}=F\circ\Phi_{t,s}\circ\Phi_s$ need not be smaller than $F_s=F\circ\Phi_s$ and $F_t$ may then increase in the time-interval $[s,t]$ (see~\cite{ChruKoss3} for an updated review).

In the following we study the discrete-time occurrence of such a non-Markovianity criterion.

\section{Witnessing non-Markovianity}

In the continuous case, several witnesses of non-Markovianiy have been proposed~\cite{PiiloManiscalco}-\cite{vas:PRA:11} based on the absence of monotonic behaviour.
In the following, we will consider the Hilbert-Schmidt norm on the space of bounded operators on $\HI$,
\begin{equation}
||\hat{X}||_{HS}=\sqrt{\Tr({\hat{X}}^{\dagger}\hat{X})}\ .
\label{SM36}
\end{equation}
All unital completely positive maps $\Psi$, that is such that $\Psi[1]=1$, satisfy the so-called Schwartz-positivity, namely
\begin{equation}
\label{SP}
\Psi[\hat{X}^\dag\,\hat{X}]\geq (\Psi[X])^\dag\Psi[X]
\end{equation}
for all operators $\hat{X}$ on $\HI$.
Indeed, $\Psi$ completely positive means that $\Psi\otimes \id_{n}$ is positive; then, setting $n=2$, on the positive operator on $\HI\otimes\CI^2$
$$
\begin{pmatrix}\hat{X}^\dag\cr-1\end{pmatrix}(\hat{X},-1)=\begin{pmatrix}
\hat{X}^{\dagger}\hat{X} &-\hat{X}^{\dagger}\cr

         -\hat{X}   &      1
\end{pmatrix}
$$
one gets
$$
\begin{pmatrix}
\Psi[\hat{X}^{\dagger}\hat{X}] &-(\Psi[\hat{X}])^{\dagger}\cr

         -\Psi[\hat{X}]  &      1
\end{pmatrix}\,\geq\,0\ .
$$
It then follows that, for all $\vert\varphi\rangle\in \HI$,
$$
\left(\langle\varphi|,\langle\varphi|(\Psi[X])^\dag\right)
\begin{pmatrix}
\Psi[\hat{X}^\dagger\hat{X}] &-\Psi[\hat{X}^\dag]\\

         -\Psi[\hat{X}]  &      1
\end{pmatrix}
\begin{pmatrix}
|\varphi\rangle\cr
\Psi[X]|\varphi\rangle
\end{pmatrix}
=
\langle\varphi\vert\Psi[\hat{X}^\dag\hat{X}]-(\Psi[\hat{X}])^\dag\Psi[\hat{X}]\vert\varphi\rangle\ge0\ .
$$
From Schwartz-positivity it follows that the Hilbert-Schmidt norm must decrease under completely positive unital and trace preserving maps; indeed,
$$
\|\Psi[\hat{X}]\|^2_{HS}=\Tr\Big((\Psi[\hat{X}])^\dag\Psi[\hat{X}]\Big)\leq\Tr\Big(\Psi[\hat{X}^\dag\hat{X}]\Big)=\Tr\Big(\hat{X}^\dag\hat{X}\Big)
=\|\hat{X}\|^2_{HS}\ .
$$
Let us now consider the Hilbert-Schmidt norm of a density matrix $\hat{\rho}_N=\Phi_N[\hat{\rho}]$ that evolves in time
according to the discrete family of dynamical maps $\Phi_N$ in~(\ref{SM32}) that compose as in~(\ref{intertwining}).
If $\Phi_{N,M}$ were completely positive, then, because of the previous argument
\begin{equation}
\label{HSmon}
||\hat{\rho}_N||^2_{HS}=||\Phi_{N,M}[\hat{\rho}_M]||^2_{HS}=\Tr\Big(\left(\Phi_{N,M}[\hat{\rho}_N]\right)^\dagger\Phi_{N,M}[\hat{\rho}_M]\Big)\leq \Tr(\Phi_{N,M}[\hat{\rho}^2_M])=\|\hat{\rho}_M||^2_{HS}\ ,
\end{equation}
for all $0\leq M\leq N$.
Indeed, $\Phi_{N,M}=\Phi_N\circ\Phi_M^{-1}$ is unital and trace preserving for such are $\Phi_N$ and $\Phi_M^{-1}$ as follows from~(\ref{unital0}),~(\ref{trace0}) and~(\ref{unital}),~(\ref{trace}).

A sufficiently explicit expression for the Hilbert-Schmidt norm $\|\hat{\rho}_N\|_{HS}$ that is amenable to numerical investigations is
readily achieved.
Using~(\ref{SM33}), one explicitly finds
\begin{eqnarray}
\nonumber
&&
\|\Phi_N[\hat{\rho}]\|^2_{HS}=\Tr\Big(\Psi^2_N[\hat{\rho}]\Big)\\
\label{app1}
&&
=\sum_{\bar{x}_N,\bar{y}_N\in\{0,1\}^N}\,p(\bar{x}_N)\,p(\bar{y}_N)\,
\Tr\left(\hat{\rho}\,e^{i\langle\bar{x}_N-\bar{y}_N\vert\bar{\nu}_N(\hat{\theta})\rangle}\,\hat{\rho}\,
e^{-i\langle\bar{x}_N-\bar{y}_N\vert\bar{\nu}_N(\hat{\theta})\rangle}\right)\ .
\end{eqnarray}
By going to the angle representation, whereby $\bar{\nu}_N(\hat{\theta})\vert\theta\rangle=\bar{\nu}_N(\theta)\vert\theta\rangle$, one expresses the Hilbert-Schmidt norm in terms of the probability distribution characteristic function:
\begin{equation}
\label{app2}
\|\Phi_N[\hat{\rho}]\|^2_{HS}=
\int_0^{2\pi}{\rm d}\theta_1\,\int_0^{2\pi}{\rm d}\theta_2\,|\langle\theta_1\vert\hat{\rho}\vert\theta_2\rangle|^2\, \Big|\Lambda\Big(\bar{\nu}_N(\theta_1)-\bar{\nu}_N(\theta_2)\Big)\Big|^2\ .
\end{equation}
In the case of the Markov process one uses~(\ref{charfunctmarkov}) with
$$
u_j=\nu_j(\theta_1)-\nu_j(\theta_2)=K\Big(\cos(\theta_1+j\tau)-\cos(\theta_2+j\tau)\Big)\ ,\quad j=1,2,\ldots,N\ .
$$

Figure $5$  shows various time behaviours of the Hilbert-Schmidt norm for an initial momentum
eigen-projector $\vert0\rangle\langle0\vert$: in absence of memory ($a=0$), the Hilbert-Schmidt distance  behaves monotonically, whereas the higher is $a$, the higher is the memory in the process, the more it oscillates.

\begin{figure}[H]
\begin{center}
\includegraphics[width=0.7\textwidth]{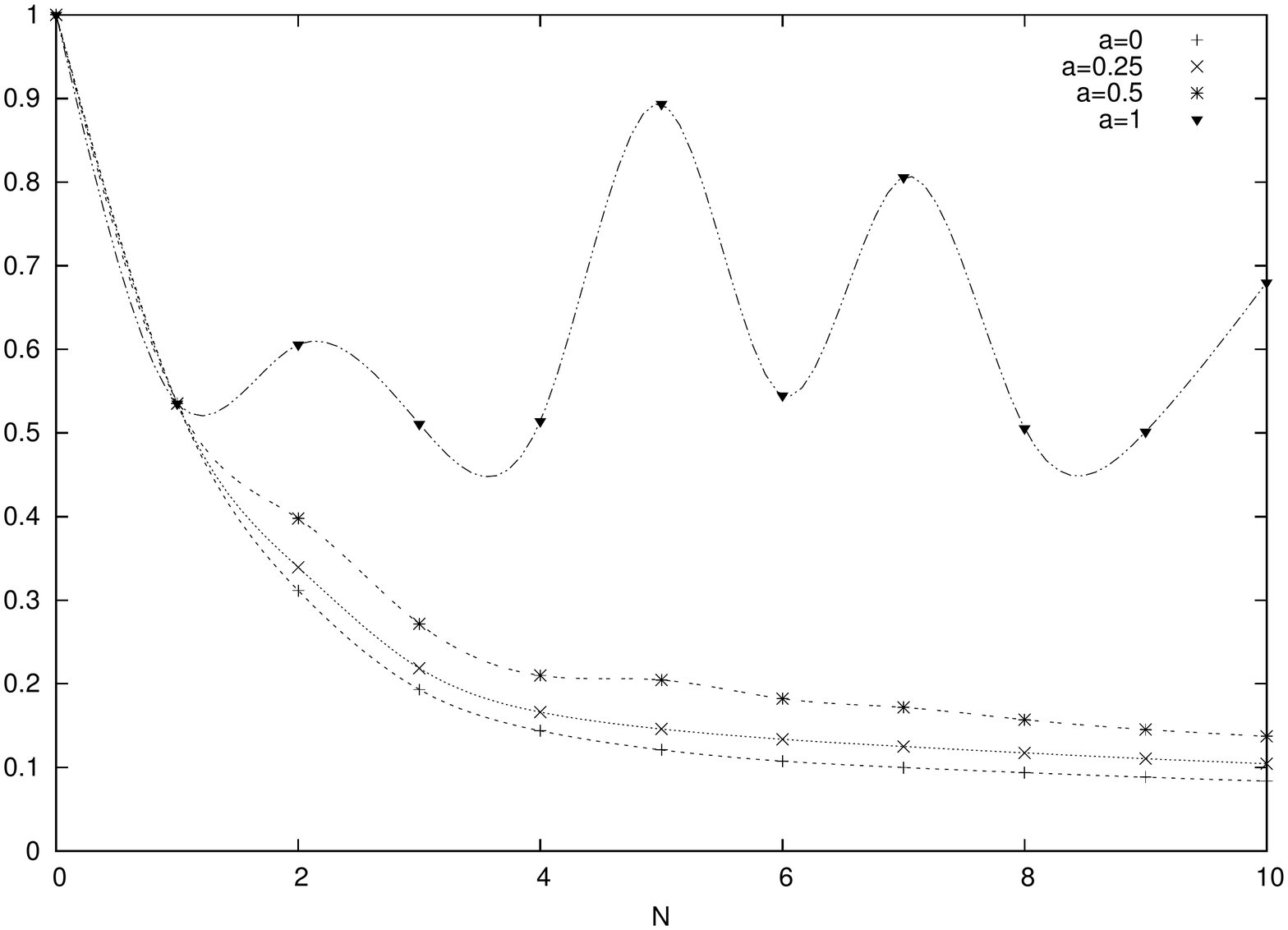}
\caption{Time behaviour of $\|\Phi_N[\vert 0\rangle\langle 0\vert]\|^2_{HS}$ for Markovian kicks: $K=3$ and $\tau=2\pi\sqrt{2}$.}
\label{IM4}
\end{center}
\end{figure}

\begin{remark}
With respect to definition of non-Markovianity corresponding to the absence of intertwining completely positive maps, the above result is an evidence of the fact that this choice makes sense also when there are no generators for the dynamics.
Moreover, in the present case the non-Markovianity witness is the Hilbert-Schmidt norm while in~\cite{vas:PRA:11} is the fidelity and in~\cite{RivasHuelgaPlenio} the relative entropy of entanglement. This is due to the occurring maps being not only trace-preserving, but also unital.
\end{remark}

\subsection{Study of the intertwining maps}

From the previous numerical evidence, some of the maps $\Phi_{N,M}$ cannot be $2$-positive, for instance $\Phi_{2,1}\otimes\id_2$ cannot be positive, and thus $\Phi_{2,1}$ cannot be completely positive.
In this section, we address the properties of $\Phi_{1,2}$ in more detail: namely, we also consider the problem of its positivity as a linear map, that is whether $\Phi_{2,1}[P]\geq 0$ on all projectors $P$ on $\HI$.
We set $\Gamma=\Phi_{2,1}=\Phi_2\circ\Phi_1^{-1}$.
By means of~(\ref{SM33}) with $N=2$, of~(\ref{inverse1}) and of the fact that
$$
e^{-i\hat{p}\ell\tau}\,\cos(\hat{\theta})\,e^{i\hat{p}\ell\tau}=\cos(\hat{\theta}-\ell\tau)\ ,
$$
with $N=1$ one finds
\begin{eqnarray}
\label{gamma1}
&&\hskip -.5cm
\Gamma[\hat{\rho}]=
\sum_{x_{1,2}\in\{0,1\}}p(x_1,x_2)\, \int{\rm d}u{\rm d}v\,\frac{e^{iuv}}{2\pi\Lambda_1(u)}\,
E(x_1,x_2,v,\hat{\theta})\,e^{-i\hat{p}\tau}\,\rho\, e^{i\hat{p}\tau}\,E^\dag(x_1,x_2,v,\hat{\theta})\\
\label{gamma2}
&&\hskip -.5cm
E(x_1,x_2,v,\hat{\theta})=e^{iK(x_{1}+v)\cos(\hat{\theta}-\tau)}e^{-iK(1-x_{2})\cos\hat\theta}.
\end{eqnarray}
In the angle representation where $\cos(\hat{\theta})\vert\theta\rangle=\cos(\theta)\vert\theta\rangle$,
\begin{eqnarray}
\label{gamma3}
&&\hskip -.5cm
\Gamma[\hat{\rho}]=
\int_0^{2\pi}{\rm d}\theta_1\int_0^{2\pi}{\rm d}\theta_2\,e^{iK(\cos\theta_2-\cos\theta_1)}\, G(\theta_1,\theta_2)\,
\vert\theta_1\rangle\langle\theta_1\vert\,e^{-i\hat{p}\tau}\,\rho\, e^{i\hat{p}\tau}\,\vert\theta_2\rangle\langle\theta_2\vert\\
&&\hskip -.5cm
G(\theta_1,\theta_2)=\frac{\Lambda_2\Big(K[\cos(\theta_2-\tau)-\cos(\theta_1-\tau)],K[\cos\theta_2-\cos\theta_1]\Big)}
{\Lambda_1\Big(K[\cos(\theta_2-\tau)-\cos(\theta_1-\tau)]\Big)}\ .
\label{gamma4}
\end{eqnarray}
Let $\hat{\rho}$ be the following pure state projection
\begin{eqnarray}
\label{purestate}
\rho
&=&e^{i\hat{p}\tau}\,e^{iK\cos\hat{\theta}}\,\vert0\rangle\langle0\vert\,
e^{-iK\cos\hat{\theta}}\,e^{-i\hat{p}\tau}\qquad\hbox{so that}\\
\label{positivity}
\Gamma[\hat{\rho}]&=&\frac{1}{2\pi}
\int_0^{2\pi}{\rm d}\theta_1\int_0^{2\pi}{\rm d}\theta_2\, G(\theta_1,\theta_2)\,
\vert\theta_1\rangle\langle\theta_2\vert .
\end{eqnarray}
Let then $\HI\ni\vert\varphi\rangle=\vert\varphi_1\rangle-\vert\varphi_2\rangle$ be a not normalized vector, with
\begin{equation}
\label{positivity1}
\langle\theta\vert\varphi_j\rangle=\frac{1}{\varepsilon}\chi_{[\bar{\theta}_j-\varepsilon/2,\bar{\theta}_j+\varepsilon/2]}(\theta)\ ,\quad j=1,2
\end{equation}
the characteristic functions of intervals around $\bar{\theta}_j$, $j=1,2$, of size $\varepsilon\geq 0$. Then one considers the mean value
\begin{eqnarray}
\label{positivity2}
\hskip -.5cm
\Delta(\varepsilon)&=&2\pi\,\langle\varphi\vert\Gamma[\hat{\rho}]\vert\varphi\rangle=I_1(\varepsilon)+I_2(\varepsilon)-I_{12}(\varepsilon)
-I_{21}(\varepsilon)\quad\hbox{where}\\
\nonumber\\
\label{positivity2a}
I_1(\varepsilon)&=&\frac{1}{\varepsilon^2}
\int_{\bar{\theta}_1-\varepsilon/2}^{\bar{\theta}_1+\varepsilon/2}{\rm d}\theta_1\int_{\bar{\theta}_1-\varepsilon/2}^{\bar{\theta}_1+\varepsilon/2}{\rm d}\theta_2\, G(\theta_1,\theta_2)\\
\nonumber\\
\label{positivity2b}
I_2(\varepsilon)&=&\frac{1}{\varepsilon^2}
\int_{\bar{\theta}_2-\varepsilon/2}^{\bar{\theta}_2+\varepsilon/2}{\rm d}\theta_1
\int_{\bar{\theta}_2-\varepsilon/2}^{\bar{\theta}_2+\varepsilon/2}{\rm d}\theta_2\, G(\theta_1,\theta_2)\\
\nonumber\\
\label{positivity2c}
I_{12}(\varepsilon)&=&
\frac{1}{\varepsilon^2}
\int_{\bar{\theta}_1-\varepsilon/2}^{\bar{\theta}_1+\varepsilon/2}{\rm d}\theta_1\int_{\bar{\theta}_2-\varepsilon/2}^{\bar{\theta}_2+\varepsilon/2}{\rm d}\theta_2\, G(\theta_1,\theta_2)\\
\nonumber\\
\label{positivity2d}
I_{21}(\varepsilon)&=&\frac{1}{\varepsilon^2}\int_{\bar{\theta}_2-\varepsilon/2}^{\bar{\theta}_2+\varepsilon/2}{\rm d}\theta_1
\int_{\bar{\theta}_1-\varepsilon/2}^{\bar{\theta}_1+\varepsilon/2}{\rm d}\theta_2\, G(\theta_1,\theta_2)\ .
\end{eqnarray}
By letting $\varepsilon\to0$ one gets
\begin{eqnarray}
\nonumber
\Delta&=&\lim_{\varepsilon\to0}\Delta(\varepsilon)=G(\bar{\theta}_1,\bar{\theta}_1)+G(\bar{\theta}_2,\bar{\theta}_2)-
G(\bar{\theta}_1,\bar{\theta}_2)-G(\bar{\theta}_2,\bar{\theta}_1)\\
\label{nonppositivity}
&=&2-\Big(G(\bar{\theta}_1,\bar{\theta}_2)+G(\bar{\theta}_2,\bar{\theta}_1)\Big)\ .
\end{eqnarray}
Using~(\ref{charfunctmarkov}), the quantity $\delta(\bar{\theta}_1,\bar{\theta}_2)$ between parenthesis reads
\begin{eqnarray}
\nonumber
&&
\delta(\bar{\theta}_1,\bar{\theta}_2)=\frac{2\cos\frac{K\big(\cos\theta_{2}-\cos\theta_{1}\big)}{2}}
{\cos\frac{K\big(\cos(\theta_{2}-\tau)-\cos(\theta_{1}-\tau)\big)}{2}}\,\times\\
\nonumber
&&\hskip .5cm \times\Bigg(
\cos\frac{K\big(\cos(\theta_{2}-\tau)-\cos(\theta_{1}-\tau)\big)}{2}\cos\frac{K\big(\cos\theta_{2}
-\cos\theta_{1}\big)}{2}\\
&&\hskip 1cm
-a\sin\frac{K\big(\cos(\theta_{2}-\tau)-\cos(\theta_{1}-\tau)\big)}{2}\sin\frac{K\big(\cos\theta_{2}
-\cos\theta_{1}\big)}{2}\Bigg)
\label{gaussiandelta}\ .
\end{eqnarray}

For a suitable choice of angles $\bar{\theta}_{1,2}$, the quantity $\delta(\bar{\theta}_1,\bar{\theta}_2)$ can become strictly larger than
$2$ as showed in Fig. $6$, the positive and negative peaks correspond to zeroes of the denominator in the above expressions.
Therefore, the quantity $\Delta$ in~(\ref{nonppositivity}) becomes negative showing that the intertwining map, besides not being completely positive, is not even a positive map. In the absence of memory, that is when $a=0$, the function $\Delta$ is instead positive, as it must be.

\begin{remark}
\label{remfinal}
The functions in~(\ref{positivity1}) tend to Dirac deltas when $\varepsilon\to0$ and thus exit from the Hilbert space $\HI$; however, because of the continuity
of the functions entering the integral defining $\Delta(\varepsilon)$ in~(\ref{positivity2}), the previous limit ensures that one can always find a suitably small, but finite $\varepsilon$ such that $\Delta(\varepsilon)<0$.
\end{remark}

\begin{figure}[H]
\begin{center}
\includegraphics[width=0.7\textwidth]{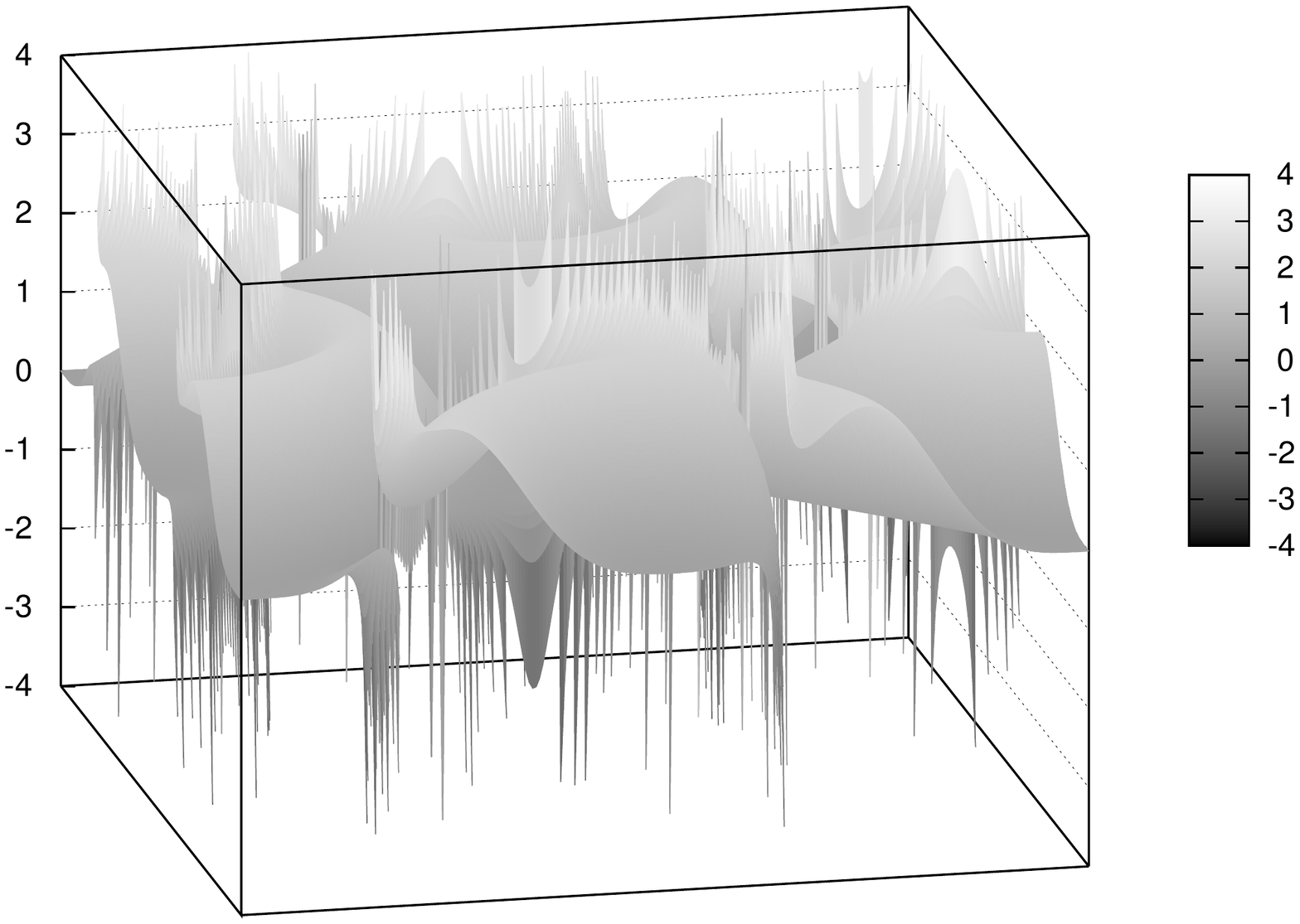}
\caption{Plot of the function $\Delta$:  $[0;2\pi]\times[0;2\pi]\to\mathbb{R}$ for Markovian kicks : $a=0.1$, $K=3$ and $\tau=2\pi\sqrt{2}$.}
\label{IM5}
\end{center}
\end{figure}

\section{Conclusions}

We studied a stochastic version of the quantum linear kicked rotor, obtained by letting the kick strengths become a discrete Markov process.
The dynamics becomes a dissipative quantum process consisting of a discrete family of completely positive maps. In case of absence of memory, when the Markov process becomes of Bernoulli type, it has been shown that these maps compose as a discrete semigroup. On the other hand,  when memory effects are present, non-Markovianity is witnessed by the non monotonic behaviour of the Hilbert-Schmidt norm. The origin of this behaviour is to be found in the fact that the
intertwining maps connecting completely positive maps at different ticks of time are trace-preserving but not completely positive, linking non-Markovianity to non-divisibility (in a discrete sense in the present case). A more detailed study of the intertwining maps shows that they are not even positive.

\end{document}